\documentclass{article}
\usepackage{graphicx,amssymb,amsfonts,amsmath,hyperref}
\begin{document}
\title{Extending Feynman's Formalisms for
Modelling Human Joint Action Coordination}
\author{Vladimir G. Ivancevic, Eugene V. Aidman and Leong Yen}\date{}\maketitle

\begin{abstract}
The recently developed Life--Space--Foam approach to goal-directed human action deals with individual actor dynamics. This paper
applies the model to characterize the dynamics of co-action by two
or more actors. This dynamics is modelled by: (i) a two-term joint
action (including cognitive/motivatonal potential and kinetic
energy), and (ii) its associated adaptive path integral,
representing an infinite--dimensional neural network. Its feedback
adaptation loop has been derived from Bernstein's concepts of
sensory corrections loop in human motor control and Brooks'
subsumption architectures in robotics. Potential applications of
the proposed model in human--robot interaction research are
discussed.\bigbreak

\end{abstract}

\section{Introduction}

Recently \cite{IA} we have suggested a generalized motivational/cognitive
action, generating Lewinian force--fields \cite{Lewin51,Lewin97} on smooth
manifolds. On the other hand, cognitive neuroscience investigations,
including fMRI studies of human co-action, suggest that cognitive and neural
processes supporting co-action include joint attention, action observation,
task sharing, and action coordination \cite{Fogassi,Knoblich,Newman,Sebanz}.
For example, when two actors are given a joint control task (e.g., tracking
a moving target on screen) and potentially conflicting controls (e.g., one
person in charge of acceleration, the other -- deceleration), their joint
performance depends on how well they can anticipate each other's actions. In
particular, better coordination is achieved when individuals receive
real-time feedback about the timing of each other's actions \cite{Sebanz}.

\section{The Action--Amplitude Model}	

To model the dynamics of the joint human action, we associate each of the actors
with an $n-$dimensional ($n$D, for short) Riemannian Life--Space manifold,
that is a set of their own time dependent trajectories, $M_{\alpha
}=\{x^{i}(t_{i})\}$ and $M_{\beta }=\{y^{j}(t_{j})\}$, respectively. Their
associated tangent bundles contain their individual $n$D (loco)motion
velocities, $TM_{\alpha }=\{\dot{x}^{i}(t_{i})=dx^{i}/dt_{i}\}$ and $%
TM_{\beta }=\{\dot{y}^{j}(t_{j})=dy^{j}/dt_{j}\}.$

Following \cite{IA}, we use the modelling machinery consisting of:

1. Adaptive joint action (\ref{Fey1})--(\ref{Fey3}) at the top--master
level, describing the externally--appearing deterministic, continuous and
smooth dynamics, ~and

2. Corresponding adaptive path integral (\ref{pathInt}) at the bottom--slave
level, describing a wildly fluctuating dynamics including both continuous
trajectories and Markov chains. This lower--level joint dynamics can be
further discretized into a partition function of the corresponding
statistical dynamics.

\subsection{Adaptive joint action}

By adapting and extending classical Wheeler--Feynman \textit{%
action--at--a--distance electrodynamics} \cite{FW} and applying it to human
co--action, we propose a two--term \emph{psycho--physical action} (summation convention is
always assumed):
\begin{eqnarray}
A[x,y;t_{i},t_{j}] &=&\frac{1}{2}\int_{t_{i}}\int_{t_{j}}\alpha _{i}\beta
_{j}\,\delta (I_{ij}^{2})\,\,\dot{x}^{i}(t_{i})\,\dot{y}^{j}(t_{j})\,%
\,dt_{i}dt_{j}+{\frac{1}{2}}\int_{t}g_{ij}\,\dot{x}^{i}(t)\dot{x}^{j}(t)\,dt
\notag \\
\text{with\qquad }I_{ij}^{2} &=&\left[ x^{i}(t_{i})-y^{j}(t_{j})\right]
^{2},\qquad \text{where \  \ }IN\leq t_{i},t_{j},t\leq OUT.\hspace{2cm}
\label{Fey1}
\end{eqnarray}

The first term in (\ref{Fey1}) represents \emph{potential energy of the
cognitive/motivational interaction} between the two agents $\alpha _{i}$ and
$\beta _{j}$.\footnote{%
Although, formally, this term contains cognitive velocities, it still
represents `potential energy' from the physical point of view.} It is a
double integral over a delta function of the square of interval $I^{2}$
between two points on the paths in their Life--Spaces; thus, interaction
occurs only when this interval, representing the motivational cognitive
distance between the two agents, vanishes. Note that the cognitive
(loco)motions of the two agents $\alpha _{i}[x^{i}(t_{i})]$ and $\beta
_{j}[y^{j}(t_{j})]$, generally occur at different times $t_{i}$ and $t_{j}$\
unless $t_{i}=t_{j},$ when \emph{cognitive synchronization} occurs.

The second term in (\ref{Fey1}) represents \emph{kinetic energy of the
physical interaction}. Namely, when the cognitive synchronization in the
first term takes place, the second term of physical kinetic energy is
activated in the common manifold, which is one of the agents' Life Spaces,
say $M_{\alpha }=\{x^{i}(t_{i})\}$.

The reason why we have chosen the action (\ref{Fey1}) as a macroscopic model for human joint action is that (\ref{Fey1}) naturally represents the transition map,
$$A[x,y;t_{i},t_{j}]~:~{\rm MENTAL~INTENTION}~\stackrel{Synch}\Longrightarrow~{\rm PHYSICAL~ACTION},$$
from mutual cognitive intention to joint physical action, in which the joint action starts after the mutual cognitive intention is synchronized. In simple words, ``we can efficiently act together only after we have tuned--up our intentions."

Similarly, if we have the joint action of three agents, say $\alpha _{i}$, $%
\beta _{j}$ and $\gamma _{k}$ (e.g., $\alpha _{i}$ in charge of
acceleration, $\beta _{j}$ -- deceleration and $\gamma _{k}-$ steering), we
can associate each of them with an $n$D Riemannian Life--Space manifold, say
$M_{\alpha }=\{x^{i}(t_{i})\}$, $M_{\beta }=\{y^{j}(t_{j})\}$, and $%
M_{\gamma }=\{z^{k}(t_{k})\},$ respectively, with the corresponding tangent
bundles containing their individual (loco)motion velocities, $TM_{\alpha }=\{%
\dot{x}^{i}(t_{i})=dx^{i}/dt_{i}\},TM_{\beta }=\{\dot{y}%
^{j}(t_{j})=dy^{j}/dt_{j}\}$ and $TM_{y}=\{\dot{z}^{k}(t_{k})=dz^{k}/dt_{k}%
\}.$ Then, instead of (\ref{Fey1}) we have
\begin{eqnarray}
A[t_{i},t_{j},t_{k};t] =\frac{1}{2}\int_{t_{i}}\int_{t_{j}}\int_{t_{k}}%
\alpha _{i}(t_{i})\beta _{j}\,(t_{j})\,\gamma _{k}\,(t_{k})\delta
(I_{ijk}^{2})\,\,\dot{x}^{i}(t_{i})\,\dot{y}^{j}(t_{j})\,\dot{z}%
^{k}(t_{k})\,dt_{i}dt_{j}dt_{k}\hspace{1cm} \notag  \label{Fey11} \\
+~{\frac{1}{2}}\int_{t}W_{rs}^{M}(t,q,\dot{q})\,\dot{q}^{r}\dot{q}%
^{s}\,dt,\,\qquad (\text{where }IN \leq t_{i},t_{j},t_{k},t\leq
OUT)\hspace{3cm}
\label{Fey2} \\
\text{with}\qquad I_{ijk}^{2}
=[x^{i}(t_{i})-y^{j}(t_{j})]^{2}+[y^{j}(t_{j})-z^{k}(t_{k})]^{2}+[z^{k}(t_{k})-x^{i}(t_{i})]^{2},
\hspace{2cm}\notag  \label{Fey3}
\end{eqnarray}

The triple joint action (\ref{Fey2})\footnote{as well as its
$N$D--generalizations} has a considerably more complicated
geometrical structure then the bilateral
co--action (\ref{Fey1}). It actually happens in the common $3n$D \textit{%
Finsler manifold} $M_{J}=M_{\alpha }\cup M_{\beta }\cup M_{y}$,
parameterized by the local joint coordinates dependent on the common time $t$%
. That is, $M_{J}=\{q^{r}(t),\,r=1,...,3n\}.$ Geometry of the joint manifold
$M_{J}$ is defined by the \textit{Finsler metric function} $%
ds=F(q^{r},dq^{r}),$ defined by
\begin{equation}
F^{2}(q,\dot{q})=g_{rs}(q,\dot{q})\dot{q}^{r}\dot{q}^{s},\qquad \text{(where
}g_{rs}\text{ is the Riemann metric tensor)}  \label{Fins1}
\end{equation}%
\ and the \textit{Finsler tensor} $C_{rst}(q,\dot{q}),$ defined by (see \cite%
{GaneshSprBig,GaneshADG})%
\begin{equation}
C_{rst}(q,\dot{q})=\frac{1}{4}\frac{\partial ^{3}F^{2}(q,\dot{q})}{\partial
\dot{q}^{r}\partial \dot{q}^{s}\partial \dot{q}^{t}}=\frac{1}{2}\frac{%
\partial g_{rs}}{\partial \dot{q}^{r}\partial \dot{q}^{s}}.  \label{Fins2}
\end{equation}%
From the Finsler definitions (\ref{Fins1})--(\ref{Fins2}), it follows that
the partial interaction manifolds, $M_{\alpha }\cup M_{\beta },$ $M_{\beta
}\cup M_{y}$ and $M_{\alpha }\cup M_{y}$ have Riemannian structures with the
corresponding interaction kinetic energies, $$T_{\alpha \beta }=\frac{1}{2}%
g_{ij}\dot{x}^{i}\dot{y}^{j},\qquad T_{\alpha \gamma }=\frac{1}{2}g_{ik}\dot{x}%
^{i}\dot{z}^{k},\qquad T_{\beta \gamma }=\frac{1}{2}g_{jk}\dot{y}^{j}\dot{z}%
^{k}.$$

\subsection{Adaptive path integral}

At the slave level, the adaptive path integral (see \cite{IA}), representing
an infinite--dimensional neural network, corresponding to the adaptive
bilateral joint action (\ref{Fey1}), reads
\begin{equation}
\langle OUT|IN\rangle :=\int\mathcal{D}[w,x,y]\, {\mathrm e}^{\mathrm i
A[x,y;t_{i},t_{j}]}, \label{pathInt}
\end{equation}
where the Lebesgue integration is performed over all continuous paths $%
x^{i}=x^{i}(t_{i})$ and $y^{j}=y^{j}(t_{j})$, while summation is performed
over all associated discrete Markov fluctuations and jumps. The symbolic
differential in the path integral (\ref{pathInt}) represents an \textit{%
adaptive path measure}, defined as a weighted product
\begin{equation}
\mathcal{D}[w,x,y]=\lim_{N\rightarrow \infty
}\prod_{s=1}^{N}w_{ij}^{s}dx^{i}dy^{j},\qquad ({i,j=1,...,n}).  \label{prod}
\end{equation}

Similarly, in case of the triple joint action, the adaptive path integral
reads,
\begin{equation}
\langle OUT|IN\rangle :=\int\mathcal{D}[w;x,y,z;q]\, {\mathrm e}^{\mathrm i
A[t_{i},t_{j},t_{k};t]}, \label{pathInt2}
\end{equation}
with the adaptive path measure defined by%
\begin{equation}
\mathcal{D}[w;x,y,z;q]=\lim_{N\rightarrow \infty
}\prod_{S=1}^{N}w_{ijkr}^{S}dx^{i}dy^{j}dz^{k}dq^{r},\qquad
(i,j,k=1,...,n;~r=1,...,3n).  \label{prod2}
\end{equation}

\section{Chaos and Bernstein--Brooks Adaptation}

From previous sections, we can see that for modelling a two--actor co--action the Riemannian geometry is sufficient. However,
it becomes insufficient for modelling the joint action of 3 or more actors, due to an \emph{intrinsic chaotic coupling} between
the individual actors. In this case we
have to use the Finsler geometry, which is a generalization of the
Riemannian one. This corresponds to the well-known fact in chaos theory that in continuous--time systems chaos cannot exist in the phase plane -- the third dimension of the system phase--space is neccessary for its existence. This also corresponds to the well-known fact of life that a trilateral (or, multilateral) relation is many times more complex then a bilateral relation. (It is so in politics, in business, in marriage, in romantic relationships, in friendship, everywhere... Physicists would say that any bilateral relation(ship) between Alice and Bob is very likely to crash if Chris comes in between, or at least it becomes much more complicated.) This is also related to Lotka--Volterra systems \cite{Wang,Meng}, other competing systems \cite{Luo} and predator--prey systems \cite{Jiao,Gan}, as well as interacting Morris--Lecar neurons \cite{Ma}.

The adaptive path integrals (\ref{pathInt}) and (\ref{pathInt2}) incorporate
the \textit{local Bernstein adaptation process} \cite{Bernstein,Bernstein2}
according to Bernstein's discriminator concept
\begin{equation*}
desired\;state~SW(t+1)\;=\;current\;state~IW(t)\;+\;adjustment~step~\Delta
W(t).
\end{equation*}
The robustness of biological motor control systems in handling excess
degrees of freedom has been attributed to a combination of tight
hierarchical central planning and multiple levels of sensory feedback--based
self--regulation that are relatively autonomous in their operation \cite%
{Bernstein3}. These two processes are connected through a top--down process
of action script delegation and bottom--up emergency escalation mechanisms.
There is a complex interplay between the continuous sensory feedback and
motion/action planning to achieve effective operation in uncertain
environments (such as movement on uneven terrain cluttered with obstacles). In case of three or more actors, the multilateral feedback/planning loop has the purpose of \emph{chaos control} \cite{OGY,StrAttr}.

Complementing Bernstein's motor/chaos control principles is Brooks' concept of
computational \textit{subsumption architectures} \cite%
{BrooksLayered,BrooksElephants}, which provides a method for structuring
reactive systems from the bottom up using layered sets of behaviors. Each
layer implements a particular goal of the agent, which subsumes that of the
underlying layers.

For example, a robot's lowest layer could be ``avoid an object", on top of
it would be the layer ``wander around", which in turn lies under ``explore
the world". The top layer in such a case could represent the ultimate goal
of ``creating a map". In this configuration, the lowest layers can work as
fast-responding mechanisms (i.e., reflexes), while the higher layers can
control the main direction to be taken in order to achieve a more abstract
goal.

The substrate for this architecture comprises a network of finite state
machines augmented with timing elements. A subsumption compiler compiles
\emph{augmented finite state machine} descriptions into a special-purpose
scheduler to simulate parallelism and a set of finite state machine
simulation routines. The resulting networked behavior function can be
described conceptually as:
\begin{equation*}
final\;state~w(t+1)\; =\; current\;state~w(t)\;+\; adjustment~behavior
~f(\Delta w(t)).
\end{equation*}

The Bernstein \textit{weights}, or \emph{Brooks nodes}, $w^s_{ij}=w^s_{ij}(t)
$ in (\ref{prod}) are updated by the \emph{Bernstein loop} during the joint
transition process, according to one of the two standard neural learning
schemes, in which the micro--time level is traversed in discrete steps,
i.e., if $t=t_0,t_1,...,t_s$ then $t+1=t_1,t_2,...,t_{s+1}$:

\begin{enumerate}
\item A \textit{self--organized}, \textit{unsupervised}  (e.g.,
Hebbian--like \cite{Hebb}) learning rule:
\begin{equation}
w^s_{ij}(t+1)=w^s_{ij}(t)+ \frac{\sigma}{\eta}%
(w^{s,d}_{ij}(t)-w^{s,a}_{ij}(t)),  \label{Hebb}
\end{equation}
where $\sigma=\sigma(t),\,\eta=\eta(t)$ denote \textit{signal} and \textit{%
noise}, respectively, while new superscripts $d$ and $a$ denote \textit{%
desired} and \textit{achieved} micro--states, respectively; or

\item A certain form of a \textit{supervised gradient descent learning}:
\begin{equation}
w^s_{ij}(t+1)\,=\,w^s_{ij}(t)-\eta \nabla J(t),  \label{gradient}
\end{equation}
where $\eta $ is a small constant, called the \textit{step size}, or the
\textit{learning rate,} and $\nabla J(n)$ denotes the gradient of the
`performance hyper--surface' at the $t-$th iteration.
\end{enumerate}

Both Hebbian and supervised learning\footnote{%
Note that we could also use a reward--based, \textit{reinforcement learning}
rule \cite{SB}, in which system learns its \textit{optimal policy}:
\begin{equation*}
innovation(t)=|reward(t)-penalty(t)|.
\end{equation*}%
} are used in local decision making processes, e.g., at the intention
formation phase (see \cite{IA}). Overall, the model presents a set of
formalisms to represent time-critical aspects of collective performance in
tactical teams. Its applications include hypotheses generation for real and
virtual experiments on team performance, both in human teams (e.g.,
emergency crews) and hybrid human-machine teams (e.g., human-robotic crews).
It is of particular value to the latter, as the increasing autonomy of
robotic platforms poses non-trivial challenges, not only for the design of
their operator interfaces, but also for the design of the teams themselves
and their concept of operations.

\section{Conclusion}

In this paper we have applied the previously developed Life Space Foam approach
to model the dynamics of co-action by two
or more agents. This dynamics is modelled by:
\begin{enumerate}
\item a two-term adaptive joint
action, including mental cognitive/motivatonal potential and physical kinetic energy, ~and
\item its associated adaptive path integral,
representing an infinite--dimensional neural network.
\end{enumerate}
Its feedback
adaptation loop has been derived from Bernstein's concepts of
sensory corrections loop in human motor control and Brooks'
subsumption architectures in robotics. The presented model demonstrates that in case of trilateral or multilateral joint action we have the strong possibility of chaotic behavior. Potential applications of the proposed model in human--robot interaction research are discussed.

\end{document}